\documentclass[11pt,usenames,dvipsnames]{article}
\usepackage{authblk}
\usepackage{amssymb,amsmath}       
\usepackage[utf8]{inputenc}
\usepackage{gensymb}
\usepackage{feynmf}
\usepackage[numbers]{natbib}
\usepackage[T1]{fontenc}
\usepackage[nottoc]{tocbibind}
\usepackage{orcidlink}    
\usepackage{bbm}                                                       
\usepackage{tensor}                                                    
\usepackage{color}                                
\usepackage{setspace}                                                  
\usepackage{hyperref}                                                  
\usepackage[left= 1in,right=1in,top=1in,bottom=1in]{geometry}          
\usepackage{comment}                                                   
\usepackage{authblk}                                                   
\usepackage{graphicx}                                                  
\usepackage{caption}                                                    
\usepackage{subcaption}
\usepackage{enumerate}
\usepackage{float}
\usepackage{epigraph}
\usepackage{tabularx,booktabs}
\usepackage{ amssymb }
\usepackage{bbold}
\usepackage{slashed}

\usepackage{physics}
\usepackage{tikz}
\usepackage[compat=1.1.0]{tikz-feynman}
\usepackage{braket}
\numberwithin{equation}{section}

\hypersetup{                 
    colorlinks=true,         
    linkcolor=blue,          
    citecolor=red,           
    urlcolor=Violet          
}

\newcommand{\kielce}{Institute of Physics, Jan Kochanowski University, ul. Uniwersytecka 7, 25-406, Kielce, Poland}

\newcommand{\frankfurt}{Institute for Theoretical Physics, J. W. Goethe University\\
Max-von-Laue-Straße 1, 60438 Frankfurt am Main, Germany}

\title{On the failure of the $12\%$ rule in $J/\psi$ and $\psi(2S)$ decays
}
\date{}
\author[1]{Arthur Vereijken \orcidlink{0000-0001-8577-9700} \footnote{
 \href{mailto:arthur.vereijken@gmail.com}{arthur.vereijken@gmail.com}}}

\author[1,2]{Francesco Giacosa \orcidlink{0000-0002-7290-9366} \footnote{
 \href{mailto:francesco.giacosa@ujk.edu.pl}{francesco.giacosa@gmail.com}}} 
\affil[1]{\kielce}
\affil[2]{\frankfurt}

\begin{document}
\maketitle

\begin{abstract}
The $12\%$ rule relates hadronic decays of the $J/\psi$ and
$\psi(2S)$ through the ratio of their wave functions at the origin, but
is strongly violated in several vector--pseudoscalar ($VP$) channels,
most prominently in the $\rho\pi$ one. We study these decays within
an effective chiral framework including strong and electromagnetic
interactions, with the latter implemented through vector meson dominance.
The $\psi(2S)\to VP$ data are described well by a relatively minimal
model with three free parameters, yielding
$\chi^2_{\rm red}\simeq 6.71/(9-3)=1.12$. In contrast, the corresponding description
of the $J/\psi$ decays is insufficient ($\chi^2{\rm _{red}}\simeq 132/(11-3)=16.5$) and requires additional
$\mathrm{SU}(3)$ flavor breaking and electromagnetic contributions for a fair agreement with data: $\chi^2_{\rm red}\simeq 9.87/(11-6)= 1.97$. In the past, a small mixing of $J/\psi$ with an invisible vector glueball was considered as a way to describe the decay patterns. However, we will show that 
our extended fits are `de facto' equivalent to those where a vector glueball mixes with the $J/\psi$, so that we cannot distinguish whether or not the vector glueball is present. Our
results suggest that an important part of the $\rho\pi$ puzzle may arise
from additional contributions to the $J/\psi\to VP$ amplitudes rather
than from a suppression mechanism specific to the $\psi(2S)$.
\end{abstract}

\section{Introduction}

The decay of the $J/\psi$ ($n^{2S+1}L_J = 1^{3}S_{1}$) and $\psi(2S)$
($n^{2S+1}L_J = 2^{3}S_{1}$) is primarily (excluding the $\psi(2S)$ decaying
into $J/\psi$ plus something else) through a 3-gluon channel. In the
nonrelativistic limit, this amplitude is proportional to the wave function at
the origin. Denoting either charmonium state by $C=J/\psi,\psi(2S)$, the decay
width into a light hadronic final state $h$ through the 3-gluon channel is
\cite{Appelquist:1974zd,Appelquist:1975ya}
\begin{equation}\label{waveft 0}
    \Gamma(C\to h) =
    |\psi_C(\mathbf{r} = 0)|^2
    |\mathcal{M}_{C\to h}|^2.
\end{equation}
Here $\psi_C(\mathbf{r})$ is the position wavefunction of the charmonium state
$C$, while $\mathcal{M}_{C\to h}$ is the matrix element associated with the
3-gluon channel into the specific final state $h$. The matrix element also
includes phase-space corrections which depend on the mass of the $J/\psi$ or
$\psi(2S)$. If we ignore the mass difference between the two charmonium states
for now, we look at the ratio of their decay widths into a specific state.
Only the unknown ratio of wave functions remains. Computing these wave
functions is no trivial task, but without computation it is already evident
that the ratio of decay widths is almost independent of the final state $h$
into which it decays. The same argument holds for the decay into
$e^{+}e^{-}$, for which this ratio is experimentally well-known. Using PDG
data on the $e^{+}e^{-}$ channel we compute this ratio:
\begin{equation}\label{13 rule gamma}
    \frac{\Gamma(\psi(2S) \to h)}
         {\Gamma(J/\psi \to h)}
    \approx
    \frac{\Gamma(\psi(3686) \to e^{+}e^{-})}
         {\Gamma(J/\psi \to e^{+}e^{-})}
    = 42.2\%.
\end{equation}
Usually, branching ratios are more accessible experimentally. Formulating this
in terms of branching ratios leads us to the $13\%$ rule (historically often
called the $12\%$ rule, based on the older value):
\begin{equation}
    Q_{h} =
    \frac{\mathcal{B}(\psi(2S) \to h)}
         {\mathcal{B}(J/\psi \to h)}
    \approx
    \frac{\mathcal{B}(\psi(2S) \to e^{+}e^{-})}
         {\mathcal{B}(J/\psi \to e^{+}e^{-})}
    = 13.3\%.
    \label{13 rule}
\end{equation}
The 13\% rule can also be derived to include the phase space corrections we
ignored earlier. We will include this later on, but for a qualitative
description it is not needed. Most measured baryon-antibaryon decay channels
are compatible with this rule. Some meson decay channels follow this rule, but
some channels severely violate it. Including phase space corrections does not
change this picture. The $\rho \pi$ channel is the most prominent channel in
which this happens. Hence, the name given to this unexplained behavior is
``the $\rho \pi$ puzzle''. In Table \ref{tab: rho pi puzzle 1} the
experimentally measured branching ratios and $Q_h$ values for the $VP$
channels are shown. Taking into account the uncertainties, all channels are
consistent with $Q_{h}$ being suppressed relative to the $13\%$ rule.

\begin{table}[]
 \centering
    \begin{tabular}{|c|l|l|l|}
    \hline
    Decay channel $h$ & $J/\psi$ BR ($10^{-3}$) &
    $\psi(2S)$ BR ($10^{-5}$)  & $Q_{h} (\%)$   \\
    \hline
$K^0 \bar{K}^{0*} $ + c.c & $4.2\pm 0.4$ & $10.9\pm 2.0 $ &
$2.6\pm 0.5 $ \\
        $K^+ K^{-*} $ + c.c & $6.0^{+0.8}_{-1.0} $ &
        $2.9 \pm 0.4 $ & $0.48^{+0.10}_{-0.09}  $ \\
         $ \omega \eta $ & $1.74\pm 0.20 $ & $ <1.1 $ & $< 0.63$ \\
          $ \omega \eta' $ & $0.189\pm 0.018 $ &
          $3.2^{+2.5}_{-2.1}  $ & $17 ^{+13} _{- 11} $ \\
            $\phi \eta $ & $0.74 \pm 0.06$ & $3.10 \pm 0.31  $ &
            $4.2\pm 0.5 $ \\
            $ \phi \eta' $ & $0.46 \pm 0.05$ & $1.54\pm 0.20 $ &
            $3.3\pm 0.6$ \\
            $\rho \pi$ & $18.8 \pm 1.2$ & $3.2 \pm 1.2  $ &
            $0.17\pm 0.06 $ \\
            $\rho^{0} \pi^0$ & $6.2 \pm 0.6 $ & -- & -- \\
            $ \rho \eta$ & $0.193 \pm 0.023$ & $2.2 \pm 0.6 $ &
            $11.4\pm 3.4 $ \\
            $\rho \eta'$ & $0.081 \pm 0.008$ &
            $1.9^{+1.7}_{-1.2}$ & $23^{+21}_{-15}$ \\
            $\omega \pi^{0}$ & $0.45 \pm 0.05$ & $2.1 \pm 0.6 $ &
            $4.7\pm 1.4 $ \\
            \hline
    \end{tabular}
\caption{$J/\psi$ and $\psi(2S)$ branching ratios and $Q_{h}$ values for
various $VP$ decay channels from PDG \cite{ParticleDataGroup:2024cfk}.
Taking into account the errors, every channel is consistent with being
suppressed compared to the $13\%$ rule.}
    \label{tab: rho pi puzzle 1}
\end{table}

Experimentally, the first indication of the $\rho\pi$ puzzle was observed in
1983 by the Mark II collaboration \cite{Franklin:1983ve}, where the decay of
the $J/\psi$ into $\rho\pi$ was substantial, but the decay width for the
channel $\psi(2S)\to \rho \pi$ was so small it was only observed decades
later by CLEO and BES \cite{CLEO:2004vxx,BES:2004hpb}. Early on it was
noticed that the charmonium decays into $\rho\pi$ break hadron helicity
conservation \cite{Brodsky:1981kj}, which follows from perturbative QCD
predictions. Understanding why this rule is broken may have important
consequences for the applicability of perturbative QCD to the decay of
charmonium states or even other states of a similar mass.

There are many proposed explanations for the $\rho \pi$ puzzle
\cite{QuarkoniumWorkingGroup:2004kpm,Mo:2006cy}. A non-exhaustive list of
explanations includes
\begin{itemize}
\item Mixing of the $J/\psi$ with a (narrow) vector glueball \cite{Brodsky:1987bb,Hou:1982kh}
    \item Intrinsic charm components in the light mesons such as the $\rho$,
    which allow for new diagrams for the decay of the $J/\psi$, which are
    suppressed for the $\psi(2S)$ because the excited state is orthogonal to
    the charm-anticharm component of the $\rho$ \cite{Brodsky:1997fj}.

    \item Suppression of the charm-anticharm wave function at the origin for a
    component of the $\psi(2S)$ in which the $c\bar{c}$ is in a color-octet
    $^3S_1$ state \cite{Chen:1998ma}.

    \item Interference between 3-gluon and electromagnetic decays for the
    $J/\psi$ decays, while the $\psi(2S)$ decays not through a 3-gluon
    channel, but through a specific configuration of 5 gluons
    \cite{Gerard:1999uf}.

    \item A large nonvalence gluon or quark-antiquark component present in the
    wavefunction of the $\psi(2S)$, which is not significant in the $J/\psi$
    since it is the lowest lying state. The valence and nonvalence components
    then interfere destructively in the $\rho\pi$ channel
    \cite{Chernyak:1983ej}.

    \item Destructive interference between S-wave and D-wave components of the
    $\psi(2S)$, which come from mixing with the D-wave charmonium state
    $\psi''$ \cite{Rosner:2001nm}.

    \item Final state interactions, which can be of similar order as tree
    level amplitudes. In particular, $J/\psi \to \rho\pi$ would be enhanced
    by the $a_{2}\rho$ loop diagram, while for the $\psi(2S)$ this can be
    canceled by the $a_{1}\rho$ loop diagram \cite{Li:1996yn}.

    \item Cancellation between the amplitudes for the processes
    $e^{+}e^{-} \to \psi(2S)\to\rho\pi$ and the direct decay channel
    $e^{+}e^{-}\to \rho\pi$ \cite{QuarkoniumWorkingGroup:2004kpm}.
\end{itemize}

Any such explanation tends to either postulate an enhancement for the decay
$J/\psi \to \rho\pi$, or a suppression for the decay
$\psi(2S)\to \rho\pi$. In this work, rather than picking one of the options,
we will decide by the necessity to describe data. We will describe the strong
and electromagnetic decays of the $J/\psi$ and $\psi(2S)$ into $PV$ using
the framework of the extended Linear Sigma Model (eLSM) together with an
implementation of vector meson dominance (VMD) for the electromagnetic
interactions. We will first see how well the data is described without any
further considerations. If the data is not
described well in this minimal model, we extend it with other dynamics. We will find that the $\psi(2S)$ decays are described
well by just the strong and electromagnetic decays, but the $J/\psi$ needs
more ingredients. We will consider effects such as relaxing the VMD constraints, an anomalous decay channel with VMD, and an extra $\mathrm{SU}(3)$
flavor breaking contribution.

\section{Model setup}
\noindent
We assume that, like the vector glueball, the $J/\psi$ and $\psi(2S)$ are flavor blind with respect to the three light flavors, hence the form of the interaction terms is the same for all three states. The Lagrangian for decays into $PV$ is constructed using chiral symmetry, charge conjugation and parity, but dilatation invariance is broken. The leading-order strong decay into $PV$ is described by the Lagrangian \cite{Giacosa:2016hrm}
\begin{align}
\mathcal{L}_{1 \text{strong}}
&=
\frac{1}{2}g_{1V_{1}}\,
\widetilde{V}_{1,\mu\nu}\text{Tr}[L^{\mu}\Phi R^{\nu}\Phi^{\dagger}]
\nonumber\\
&\supset
g_{1V_{1}}\,
\widetilde{V}_{1,\mu\nu}
\text{Tr}[\partial^{\mu}\mathcal{P}\Phi_{0}V^{\nu}\Phi_{0}],
\label{VP lagrangian}
\end{align}
where $\widetilde{V}_{1,\mu\nu}$ is the dual field strength tensor for the initial vector state $V_{1}$, which can be either $O'$, $J/\psi$, or $\psi(2S)$. In terms of the light-flavor indices, each of the $V_{1}$ states is proportional to the identity matrix and can therefore be taken out of the traces. The fields
$R^{\mu}=V^{\mu}-A_{1}^{\mu}$ and
$L^{\mu}=V^{\mu}+A_{1}^{\mu}$ contain the vector and axial-vector nonets $V$ and $A_{1}$. Spontaneous chiral symmetry breaking induces a mixing of the axial-vector and pseudoscalar fields, incorporated through the shift
$A_{1\,\mu}\rightarrow A_{1\,\mu}+\partial_{\mu}\mathcal{P}$, where $\mathcal{P}$ is a renormalized pseudoscalar nonet. The nonet $\Phi=S+iP$ contains scalar and pseudoscalar mesons, but for the decays considered here only its condensate $\Phi_{0}$ is relevant. The exact forms of the mesonic nonets are given in Appendix \ref{appendix:nonets}.

We also include the higher-order Lagrangian with multiple traces, which is subleading in the large-$N_{c}$ limit:
\begin{align}
\mathcal{L}_{2 \text{strong}}
&=
\frac{1}{2}g_{2V_{1}}\,
\widetilde{V}_{1,\mu\nu}
\text{Tr}[L^{\mu}]
\text{Tr}[R^{\nu}]
\text{Tr}[\Phi\Phi^{\dagger}]
\nonumber\\
&\supset
g_{2V_{1}}\,
\widetilde{V}_{1,\mu\nu}
\text{Tr}[\partial^{\mu}\mathcal{P}]
\text{Tr}[V^{\nu}]
\text{Tr}[\Phi_{0}\Phi_{0}].
\label{Lag strong2}
\end{align}
The interaction in \eqref{VP lagrangian} already corresponds to a disconnected and hence OZI-suppressed diagram, while the interaction in \eqref{Lag strong2} is associated with a doubly disconnected diagram \cite{Escribano:2009jti}.

To include the electromagnetic interactions, we use a vector meson dominance model \cite{OConnell:1995nse}. The kinetic term of the initial vector field contains a mixing with the photon governed by the coupling $g_{V_{1}\gamma}$:
\begin{equation}
-\frac{1}{4}V_{1\mu\nu}V_{1}^{\mu\nu}
\to
-\frac{1}{4}V_{1\mu\nu}V_{1}^{\mu\nu}
-\frac{g_{V_{1}\gamma}}{2}V_{1\mu\nu}F^{\mu\nu}
+\cdots .
\label{eq V1 photon mixing}
\end{equation}
In momentum space,
$V_{1,\mu\nu}F^{\mu\nu}\to2q^{2}V_{1,\mu}A^{\mu}$.
The factor $q^{2}$ cancels against the photon propagator, while the factor 2 cancels the $1/2$, so that a $V_{1}$-photon transition contributes a factor $-g_{V_{1}\gamma}$.

To couple the photon to light mesons, we assume one universal coupling for all light vector mesons and implement it, as in Ref.~\cite{Jafarzade:2022uqo}, through the shift
\begin{equation}
V_{\mu\nu}\rightarrow V_{\mu\nu}+\frac{e}{g_{\rho}}F_{\mu\nu}Q,
\label{eq V photon mixing}
\end{equation}
where $e$ is the electromagnetic charge and $g_{\rho}\approx5.0$ \cite{OConnell:1995nse} is a universal coupling of the light vector mesons to the photon, and $Q=\text{diag}(2/3,-1/3,-1/3)$ is the charge matrix. As for the charmonium states, a vertex in which a photon turns into a vector meson, or vice versa, is associated with a factor $-\frac{e}{g_{\rho}}Q$. At the level of the fields, the shift can be implemented simply as
\begin{equation}
V_{\mu}\to V_{\mu}+\frac{e}{g_{\rho}}A_{\mu}Q.
\label{eq: shiftvmd field}
\end{equation}

We also need to couple the pseudoscalar mesons to the photon. To this end, we first introduce a flavor-invariant Lagrangian for a $PVV$ interaction,
\begin{equation}
\mathcal{L}_{PVV}
=
g_{PVV}\text{Tr}[PV_{\mu\nu}\widetilde{V}^{\mu\nu}].
\end{equation}
This interaction can also be obtained from a chirally invariant Lagrangian and initially contains only kinematically forbidden decays. Performing the shift \eqref{eq V photon mixing} generates the interactions with the photon relevant for the present analysis:
\begin{align}
\mathcal{L}_{PVV}
\to\,
\mathcal{L}_{PVV}
&+
\left(\frac{e}{g_{\rho}}\right)^{2}
g_{PVV}
F_{\mu\nu}\widetilde{F}^{\mu\nu}
\text{Tr}[PQQ]
\nonumber\\
&+
\frac{e}{g_{\rho}}
g_{PVV}
\widetilde{F}_{\mu\nu}
\text{Tr}[P\{Q,V^{\mu\nu}\}].
\label{PVV photon}
\end{align}

\begin{figure}[h]
\includegraphics[width=1.1\linewidth]{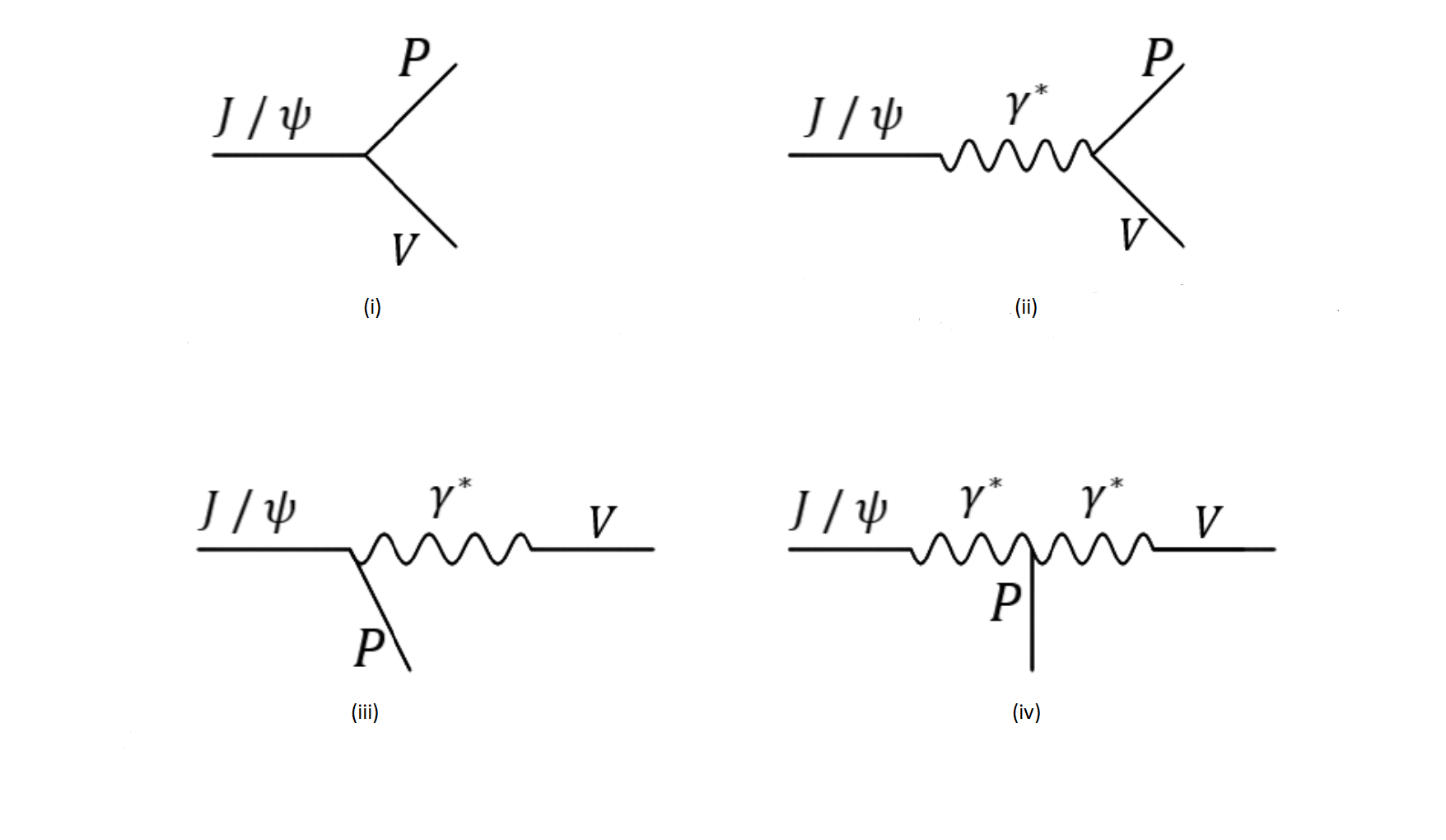}
\caption{Four Feynman diagrams for the decay channels of the $J/\psi$. Diagram (i) is a strong decay, while diagrams (ii)--(iv) proceed through electromagnetic interactions.}
\label{fig:feynmandiags}
\end{figure}

We now construct the amplitudes for the decays of the initial vector states, including the electromagnetic contributions. The Feynman diagrams for the $J/\psi$ are shown in Fig.~\ref{fig:feynmandiags}. The same diagrams apply to the $\psi(2S)$. Each diagram has the same Lorentz and phase-space structure, but different coefficients due to the couplings and flavor structure, and the different contributions can interfere constructively or destructively. After selecting a definite $PV$ final state and factoring out the common Lorentz structure, each diagram is characterized by a scalar coefficient $w_i$. The total decay width is then
\begin{align}
\Gamma_{VP}
=
\left|\sum_i w_i\right|^2
\frac{|\vec{k}|^3}{12\pi}.
\end{align}
The flavor dependence of the different contributions will be encoded in coefficients $C_i$, obtained by inserting the appropriate pseudoscalar and vector fields into the corresponding traces. We now discuss the four contributions shown in Fig.~\ref{fig:feynmandiags}.

\begin{enumerate}[(i)]

\item The strong decay channel is described by Eqs.~\eqref{VP lagrangian} and \eqref{Lag strong2}. After extracting the common Lorentz structure, the two strong contributions are
\begin{align}
\mathcal{L}_{1\text{strong}}
&\supset
g_{1V_{1}}\,
\widetilde{V}_{1,\mu\nu}
\text{Tr}[\partial^{\mu}\mathcal{P}\Phi_{0}V^{\nu}\Phi_{0}]
\nonumber\\
&\Longrightarrow
w_{1\text{strong}}
=
g_{1V_{1}}\,C_{1\text{strong}},
\label{wstrong1}
\\
\mathcal{L}_{2\text{strong}}
&\supset
g_{2V_{1}}\,
\widetilde{V}_{1,\mu\nu}
\text{Tr}[\partial^{\mu}\mathcal{P}]
\text{Tr}[V^{\nu}]
\text{Tr}[\Phi_{0}\Phi_{0}]
\nonumber\\
&\Longrightarrow
w_{2\text{strong}}
=
g_{2V_{1}}\,C_{2\text{strong}}.
\end{align}
Here $C_{1\text{strong}}$ and $C_{2\text{strong}}$ denote the numerical flavor coefficients obtained for the specific decay channel from the corresponding traces.

\item The second contribution follows from the initial vector $V_{1}$ converting into a virtual photon through the transition given in \eqref{eq V1 photon mixing}, followed by the virtual photon decaying into a $PV$ pair through the last term in \eqref{PVV photon}. This gives
\begin{align}
w_{\text{em1}}
=
-g_{V_{1}\gamma}g_{PVV}
\frac{e}{g_{\rho}}\,
C_{\text{em1}},
\end{align}
where $C_{\text{em1}}$ is obtained from the flavor structure
$\text{Tr}[P\{Q,V\}]$.

\item The third contribution comes from shifting $V$ according to \eqref{eq: shiftvmd field} in the $V_{1}PV$ vertex of \eqref{VP lagrangian}, followed by the virtual photon transitioning into a vector meson through \eqref{eq V photon mixing}. Note that the doubly disconnected vertex \eqref{Lag strong2} does not contribute through the shift \eqref{eq V photon mixing}, since $\text{Tr}[Q]=0$. The corresponding coefficient is
\begin{align}
w_{\text{em2}}
=
-g_{1V_{1}}
\left(\frac{e}{g_{\rho}}\right)^2
C_{\text{em2}},
\end{align}
where $C_{\text{em2}}$ follows from the flavor structure
$\text{Tr}[\mathcal{P}\Phi_{0}Q\Phi_{0}]\text{Tr}[VQ]$.
Similarly to \eqref{Lag strong2}, this contribution has a multiple-trace structure and is therefore expected to be suppressed compared to (ii). Furthermore, it contains a higher power of the small electromagnetic coupling.

\item The fourth contribution is generated by the initial $V_{1}$ transitioning into a virtual photon through \eqref{eq V1 photon mixing}. The virtual photon emits a pseudoscalar through the second term in \eqref{PVV photon} and subsequently transitions into a light vector meson through \eqref{eq V photon mixing}. The corresponding coefficient is
\begin{align}
w_{\text{em3}}
=
g_{V_{1}\gamma}g_{PVV}
\left(\frac{e}{g_{\rho}}\right)^3
C_{\text{em3}},
\end{align}
where $C_{\text{em3}}$ follows from the flavor structure
$\text{Tr}[PQQ]\text{Tr}[VQ]$.
This contribution also has a double-trace structure and contains a higher power of the small electromagnetic coupling, in particular the combination
$\left(\frac{e}{g_{\rho}}\right)^3$, and is therefore expected to be more strongly suppressed.

\end{enumerate}

These constitute the basic interactions of the model. We first analyze the decays of the $\psi(2S)$, for which they will be sufficient to describe the experimental data. The decays of the $J/\psi$, however, will require additional effects, such as another free coupling, an $\mathrm{SU}(3)$ flavor symmetry breaking contribution, or further relaxing of VMD universality.

\section{Results}
\subsection{Decays of the $\psi(2S)$}
\begin{figure}
    \centering
    \includegraphics[width=\linewidth]{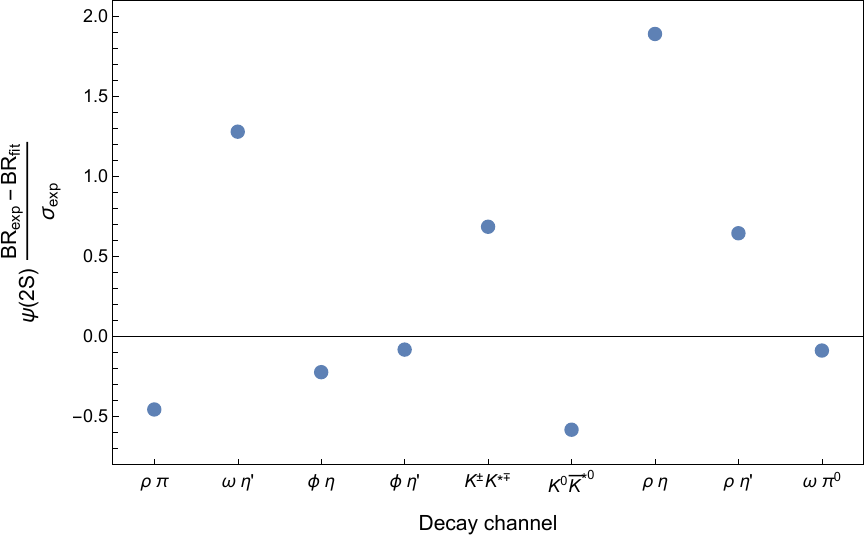}
    \caption{Pulls $\frac{\text{BR}_\text{exp}-\text{BR}_{\text{fit}}}{\sigma_{\text{exp}}}$ for the branching ratios of $\psi(2S)$ decays into $VP$ channels.}
    \label{fig:psi2s pulls}
\end{figure}

We will start with the results for the $\psi(2S)$ decays. We can reduce the number of free parameters to only those involving the $\psi(2S)$. That is, $g_{\psi},g_{2\psi},$ and $g_{\psi \gamma}$. Rather than extracting $g_{\psi \gamma}$ directly from $\Gamma(\psi(2S)\to e^{+}e^{-})$, and introducing form factors for the interaction of a highly virtual photon with light hadrons, we will instead use the VMD value $g_{\rho}\simeq 5.0$ \cite{OConnell:1995nse} and include an average of these nonperturbative effects into the fitted value of $g_{\psi\gamma}$. The coupling $g_{PVV}$ is found by matching to the $\pi^{0}\to\gamma\gamma$ decay \cite{Weinberg:1996kr}, giving
\begin{equation}
   g_{PVV}=\frac{3g_{\rho}^2}{4\pi^2 Z_{\pi}F_{\pi}} \text{ with  } Z_{\pi} = 1.709 \text{ ,}
\end{equation}
(see appendix \ref{appendix:nonets}) and $F_{\pi} = 184$ MeV being the pion decay constant.
\\
The experimental values listed in Table \ref{tab: rho pi puzzle 1} are then fitted by a standard $\chi^2$ procedure, with asymmetrical errors being symmetrized to the larger value. The fit gives the following values for the parameters
\begin{equation}
\begin{split}
    g_{\psi}&= (6.29\pm 0.24)\times 10^{-8}\, \text{MeV}^{-2}, \\
    g_{2\psi}&= (-3.3\pm 0.7)\times
   10^{-9}\,\text{MeV}^{-2}, \\
   g_{\psi\gamma}&= (0.67\pm 0.05) \times 10^{-3}.
   \label{eq:psi2s fit param}
\end{split}
\end{equation}
Because the strong coupling constants have a dimension of $\text{mass}^{-2}$ while $g_{\psi\gamma}$ is dimensionless, these values do not directly show the comparative contribution of the electromagnetic and strong diagrams.
The electromagnetic ones are also suppressed by factors of $g_{PVV}$ and $\frac{e}{g_{\rho}}$. Taking these and the flavor traces into account the hierarchy is, ignoring the zero weights present in some channels, roughly speaking,
\begin{equation}
w_{\text{1strong}}>w_{\text{em1}}\gtrsim w_{\text{2strong}}>w_{\text{em2}}>w_{\text{em3}},
\label{eq weight hierarchy}
\end{equation}
which is not entirely unexpected by the structure of the interactions. It is noteworthy that introducing the leading order electromagnetic decay channel seems to be more important than introducing the next-to-leading order strong decay channel. The total and the reduced $\chi^2$-values of the fit are
\begin{equation}
 \chi^2 \simeq 6.71 \text { , }   \chi^2_{\text{red}} = \frac{\chi^2}{d.o.f.} \simeq 1.12  \text{ .}
\end{equation}
 The comparison of the data and fit is presented in Table \ref{tab: psi2s decays} and the pulls are shown in Fig. \ref{fig:psi2s pulls}. Most of the datapoints show agreement between experiment and the fit, while a few, such as the $\omega \eta'$ and $\rho\eta$ channels differ by about two standard deviations.
\begin{table}[]
 \centering
    \begin{tabular}{|c|l|l|}
    \hline 
    Decay channel & Theoretical BR ($10^{-5}$)  & Experimental BR ($10^{-5}$)    \\
    \hline 
$K^0 \bar{K}^{0*} $ + c.c & $12.1\pm 0.8 $ &  $10.9\pm 2.0 $ \\
        $K^+ K^{-*} $ + c.c & $2.63 \pm 0.33 $& $2.9 \pm 0.4 $ \\
         $ \omega \eta $ & $0.49\pm 0.09 $ & $ <1.1 $ \\
          $ \omega \eta' $ & $< 0.037$ & $3.2^{+2.5}_{-2.1}  $
            \\
            $\phi \eta $& $3.17 \pm 0.26 $& $3.10 \pm 0.31  $\\
            $ \phi \eta' $ &$ 1.56 \pm 0.20$ &$1.54\pm 0.20 $  \\
            $\rho \pi$ &  $3.7 \pm 0.5$  & $3.2 \pm 1.2  $\\
            $\rho^{0} \pi^0$ &  $ 1.25 \pm 0.17 $ &  --  \\
            $ \rho \eta$ & $1.07\pm 0.16$ & $2.2 \pm 0.6 $ \\
            $\rho \eta'$ & $0.80 \pm 0.12$& $1.9^{+1.7}_{-1.2}$ 
            \\
            $\omega \pi^{0}$ &$2.15 \pm 0.33$ & $2.1 \pm 0.6 $ \\
            $ \phi\pi^{0} $ & $0.0089\pm 0.0014$ & $< 0.04$
            \\
            \hline
    \end{tabular}
\caption{Theoretical and experimental branching ratios of the $\psi(2S)$ for $VP$ decay channels}
    \label{tab: psi2s decays}
\end{table}

\noindent
For the $\omega \eta$ branching ratio, the PDG lists only an upper limit of $1.1\times 10^{-5}$. This was not included in the fit, but we can make a prediction using the parameters given in \eqref{eq:psi2s fit param}. This gives
\begin{equation}
    \mathcal{B}(\psi(2S) \to \omega \eta ) = (4.9\pm 0.9) \times 10^{-6},
    \label{eq: psi2s omega eta prediction}
\end{equation}
which is indeed below the given upper limit of the PDG. Using the experimental value for the $J/\psi$ branching ratio to $\omega \eta$, this gives a predicted value for $Q_{\omega\eta}=(0.28\pm 0.06)\%$, meaning the $\omega \eta$ channel violates the $13\%$ rule almost as much as the $\rho\pi$ channel, which has $Q_{\rho\pi}= (0.17 \pm 0.06) \%$.
\\
We also predict the $\phi\pi^{0}$ branching ratio, which was not included in the fit. The PDG $90\%$ confidence upper limit is $4.0\times 10^{-7}$. Our prediction is consistent with this upper limit and is
\begin{equation}
     \mathcal{B}(\psi(2S) \to \phi \pi^{0} ) = (8.9\pm 1.4) \times 10^{-8}.
\end{equation}
Another notable decay channel here is the $\omega \eta'$. While the best fit value of the theoretical branching ratio is $8\times 10^{-9}$ with a 1-sigma upper bound of $71\times 10^{-9}$, compatible within about $1.3\sigma$ of the experimental result due to the large errors, it is compatible with zero and four orders of magnitude smaller than the experimental value and the branching ratios of most other decay channels. Because it is so close to zero, a better upper bound can be found by propagating the Gaussian fit errors, which we report in Table \ref{tab: psi2s decays}. Since the branching ratio is proportional to a square of an amplitude linear in the parameters, it follows a scaled $\chi^{2}$ distribution with noncentrality $\lambda \simeq 0.061$, and so has a $90$th percentile that implies
\begin{equation}
    \mathcal{B}(\psi(2S)\to \omega \eta') < 3.7\times 10^{-7} \, \, \, (90\%\text{ propagated upper limit}).
\end{equation}
Based on the theoretical prediction, the model predicts that this channel will also severely break the $13\%$ rule as well, with the 1-sigma interval giving $Q_{\omega \eta'} < 0.037\%$, and the $90\%$ upper limit giving $Q_{\omega \eta'} < 0.20\% $. 
\\
We also predict the branching ratio for $\rho^{0}\pi^{0}$, which, even though it has a different contribution of electromagnetic channels compared to $\rho^{\pm}\pi^{\mp}$, is about one-third of the total $\rho\pi$ branching ratio, in line using only flavor counting.
\\
Overall, even though the $\psi(2S)$ is more limited in data compared to the $J/\psi$, a fairly minimal model is able to describe the experimental data well. This might suggest that the $\psi(2S)$ behaves as expected, and the special behavior that is causing the $\rho\pi$ puzzle instead comes from the $J/\psi$. 
\subsection{Decays of the $J/\psi$}
\begin{figure}
    \centering
    \includegraphics[width=\linewidth]{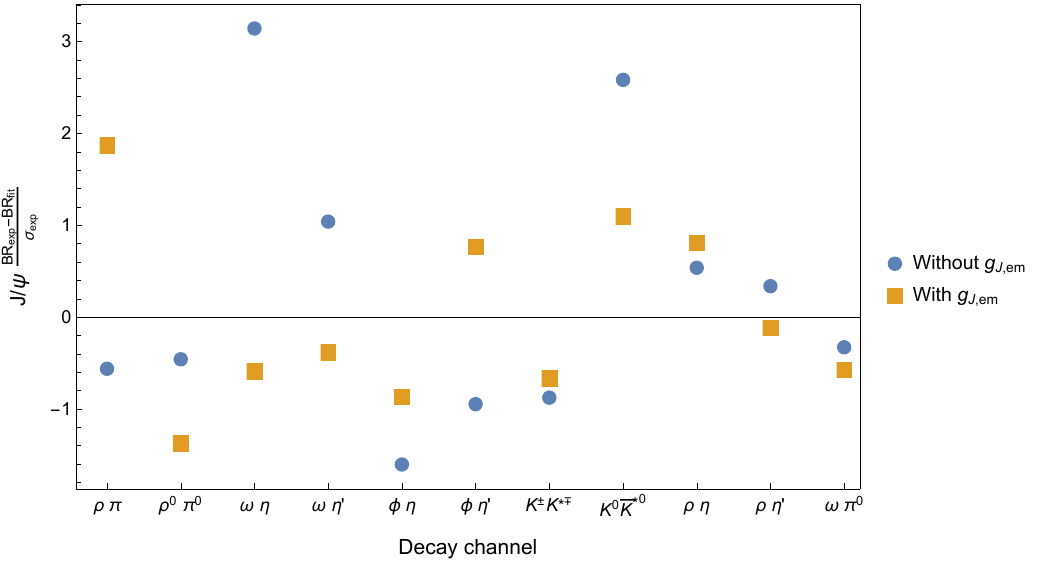}
    \caption{Pulls $\frac{\text{BR}_\text{exp}-\text{BR}_{\text{fit}}}{\sigma_{\text{exp}}}$ for $VP$ channel branching ratios of the $J/\psi$ with and without $g_{J,\mathrm{em}}$.}
    \label{fig:jpsi pulls}
\end{figure}

Next we discuss the decays of the $J/\psi$. If we use only the same interactions as we did for the $\psi(2S)$, we find a total $\chi^2 \sim 132$, meaning we are not successfully describing the data. Hence, we will need to change our description of the interactions. The first change we will make is to add an additional parameter that breaks $\mathrm{SU}(3)$ flavor symmetry in the interactions similar as in \cite{Escribano:2009jti}. At this stage, we only apply this to the leading order strong interaction, and assume that this effect on the next-to leading order terms is negligible. To this effect, we introduce the flavor breaking matrix
\begin{equation}
    X = \text{diag}(1,1,1-s),
\end{equation}
and apply it so that decay products with one $s\bar{s}$ pair have a factor $(1-\frac{1}{2}s)$ associated with it, and decay products with two $s\bar{s}$ pairs a factor $(1-s)$. The weight for the leading strong diagram then changes to
\begin{align}
     w_{1\text{strong}}&\to \frac{1}{2}g_{V_{1}} \left(\text{Tr}[\partial^{\mu}\mathcal{P}X\Phi_{0} V^{\nu} \Phi_{0}] + \text{Tr}[\partial^{\mu}\mathcal{P}\Phi_{0} V^{\nu} X \Phi_{0}] \right),
\end{align}
which for $s\to 0$ gives the previous expression \eqref{wstrong1}.
\\
For the next addition, we note that there is one more way to find an interaction corresponding to diagram (iv) in Fig. \ref{fig:feynmandiags}. Consider the vertex in which a $J/\psi$ emits a pseudoscalar:
\begin{equation}
    \mathcal{L}_{JP} \sim g'_{JP} J_{\mu\nu} \tilde{J}^{\mu\nu}\text{Tr} [P].
\end{equation}
While in itself not relevant, the $J-\gamma$ mixing \eqref{eq V1 photon mixing} lets this contribute to diagrams (iii) and (iv). It is less clear how such a term follows from a chiral Lagrangian. It turns out the consistent way to construct this interaction is via an anomalous $\mathrm{U}(1)_{A}$ breaking term of the form 
\begin{equation}
    \frac{1}{2} g'_{JP} J_{\mu\nu}\tilde{J}^{\mu\nu} i (\det\Phi - \det \Phi^{\dagger}) \supset     -g'_{JP} J_{\mu\nu}\tilde{J}^{\mu\nu} \det(\Phi_0) \text{Tr}[\Phi_{0}^{-1} P].
\end{equation}
Combined with the mixing with the photon, this will give a fourth electromagnetic contribution:
\begin{equation}
    w_{\text{em4}} = g'_{JP}(g_{J\gamma}^3-g_{J\gamma}) \det(\Phi_0) \text{Tr}[\Phi_{0}^{-1} P] \frac{e}{g_{\rho}}\text{Tr}[VQ] \equiv g_{JP} \det(\Phi_0) \text{Tr}[\Phi_{0}^{-1} P] \frac{e}{g_{\rho}}\text{Tr}[VQ],
\end{equation}
where in the last equality we have absorbed the factor $(g_{J\gamma}^3-g_{J\gamma})$ into $g_{JP}$ as they are both independent fit parameters. We will later also further relax the VMD constraints, but first we attempt to describe the data again with only these additions. We find a total $\chi^2\sim 22.9$ and a reduced $\chi^2_{\text{red}} \sim 3.8$, most of which comes from the discrepancy with the $\omega\eta$ and $K^{0}K^{*0}$ datapoints. The fit results are shown in Table \ref{tab: jpsi decays}, and the pulls in Fig. \ref{fig:jpsi pulls}. The fit results for the parameters are
\begin{equation}
\begin{split}    
    g_{J}&=  (6.83 \pm0.15 )\times 10^{-7} \text{ MeV}^{-2}             ,\\
    g_{2J}&=   (-8.91\pm0.15 )\times 10^{-8} \text{ MeV}^{-2}                ,\\
    g_{J\gamma}&=  (-2.36 \pm 0.10)\times 10^{-3}           ,\\
    g_{JP}&= (-3.4 \pm 0.7) \times 10^{-10} \text{ MeV}^{-3}              ,\\
    s&=0.510 \pm 0.019       .
\end{split}
\end{equation}
The relative contributions of the different terms are similar as in \eqref{eq weight hierarchy}, except that $w_{\text{strong}2}$ is slightly larger (but still comparable, and zero in more channels) than $w_{\text{em1}}$, and $w_{\text{em4}}$ is larger than $w_{\text{em2}}$ and $w_{\text{em3}}$.
\\
Also for the $J/\psi$ we can compute the decay into $\phi\pi^{0}$, which was not in the fit. The PDG lists two possible branching ratios for this channel, depending on the possible interference patterns \cite{BESIII:2015asx}
\begin{equation}
    \begin{split}
         \mathcal{B}(J/\psi \to \phi \pi^{0} ) = (2.94 \pm 0.16_{\text{stat}}\pm0.16_{\text{syst}} ) \times 10^{-6}, \\ \text{ or }         \mathcal{B}(J/\psi \to \phi \pi^{0} ) = (1.24\pm0.33_{\text{stat}}\pm 0.30_{\text{syst}} ) \times 10^{-7}.
    \end{split}
\end{equation}
We find
\begin{equation}
    \mathcal{B}(J/\psi \to \phi \pi^{0} ) = (2.05 \pm 0.17) \times 10^{-6},
\end{equation}
which, while there is some tension, clearly favors the first interpretation with the larger branching ratio.
\\
\\
In order to further improve the description, we introduce a new coupling $g_{J,\mathrm{em}}$ that allows the coupling in diagram (iii) to be independent of the strong coupling $g_{J}$. Explicitly, we have now for diagram (iii)
\begin{equation}
  w_{\text{em2}} = -g_{J,\mathrm{em}}
\left(\frac{e}{g_{\rho}}\right)^2
C_{\text{em2}},
\end{equation}
where $g_{J,\mathrm{em}}$ is free, while the previous fit followed the VMD prescription $g_{J,\mathrm{em}} = g_{J}$. The total and the reduced $\chi^2$-values of the fit are
\begin{equation}
 \chi^2 \simeq 9.87 \text { , }   \chi^2_{\text{red}} = \frac{\chi^2}{d.o.f.} \simeq 1.97  \text{ .}
\end{equation}
The results for the branching ratios are shown in Table \ref{tab: jpsi decays}, and the pulls in Fig. \ref{fig:jpsi pulls}. The values for the couplings are
\begin{equation}
\begin{split}    
    g_{J}&= (5.91 \pm 0.17 )\times 10^{-7} \text{ MeV}^{-2}               ,\\
    g_{J,\text{em}}&=  (-7.5 \pm 1.0 )\times 10^{-5} \text{ MeV}^{-2}                , \\
    g_{2J}&=   (-2.65 \pm 0.18 )\times 10^{-8} \text{ MeV}^{-2}              ,\\
    g_{J\gamma}&=   (-1.80\pm 0.16) \times 10^{-3}          ,\\
    g_{JP}&=  (-1.60 \pm 0.10)\times 10^{-9} \text{ MeV}^{-3}               ,\\
    s&= 0.260 \pm 0.028      .
\end{split}
\end{equation}
Here, $g_{J,\mathrm{em}}$ is substantially larger than $g_{J}$, hence we are far from the simple VMD relation $g_{J}=g_{J,\mathrm{em}}$. It is worth noting that the large fitted value of $g_{J,\mathrm{em}}$ does not translate into an equally large electromagnetic amplitude, since diagram (iii) contains the suppression factor $(e/g_\rho)^2$. Numerically,
\begin{equation*}
-g_{J,\mathrm{em}}\left(\frac{e}{g_\rho}\right)^2
\simeq 2.8\times10^{-7}\ {\rm MeV}^{-2},  
\end{equation*}
which is about $0.47\,g_J$. Thus, the fit requires an unexpectedly
sizable electromagnetic contribution, comparable to the leading
strong one. Such an enhancement could in principle reflect additional
dynamics, possibly including a hidden vector glueball component, but its
microscopic origin cannot be resolved from the present data, see also sec. \ref{sec:glueball}.
\\
The value for the $\phi\pi^{0}$ channel in this fit is 
\begin{equation}
    \mathcal{B}(J/\psi \to \phi \pi^{0} ) = (2.6 \pm 0.9) \times 10^{-5},
\end{equation}
which is about 9 times larger than the largest PDG value, but it is paired with a fairly large uncertainty. The match to the larger PDG value becomes worse, with the large uncertainty making discrimination between the two experimental datapoints more difficult, but in terms of orders of magnitude it still favors the first interpretation.
\\
Due to the number of extra parameters we have had to introduce to the decays of the $J/\psi$, the results of the $J/\psi$ fits are likely best interpreted not as a complete picture, but rather the statement that -- in stark contrast to the $\psi(2S)$ -- the $J/\psi$ data is not easily described by an effective chiral model. Contrasting the situation of the $J/\psi$ fit with the $\psi(2S)$, our results favor an interpretation in which additional dynamics enhance the $J/\psi \to \rho\pi$ (and other $VP$) decay channels, rather than one in which additional dynamics suppress the decays of the $\psi(2S)$.

\begin{table}[]
 \centering
    \begin{tabular}{|c|l|l|l|}
    \hline 
    Decay channel $h$ & Experimental BR ($10^{-3}$) & BR w/o $g_{J,\mathrm{em}}$  ($10^{-3}$)  &  BR with  $g_{J,\mathrm{em}}$  ($10^{-3}$)   \\
    \hline 
$K^0 \bar{K}^{0*} $ + c.c &$4.2\pm 0.4$ & $3.17 \pm 0.16$&  $3.76 \pm 0.18$ \\
        $K^+ K^{-*} $ + c.c & $6.0^{+0.8}_{-1.0} $& $6.88 \pm 0.25$ & $6.7 \pm 0.4  $ \\
         $ \omega \eta $ & $1.74\pm 0.20 $ & $1.11 \pm 0.06$ &$ 1.86 \pm 0.07$ \\
          $ \omega \eta' $ & $0.189\pm 0.018 $ & $0.170 \pm 0.017  $&$0.196 \pm 0.017$
            \\
            $\phi \eta $& $0.74 \pm 0.06$& $0.84 \pm 0.05  $& $0.79 \pm 0.05$ \\
            $ \phi \eta' $ &$0.46 \pm 0.05$ &$0.51 \pm 0.04$ &$0.42 \pm 0.04$ \\
            $\rho \pi$ &  $18.8 \pm 1.2$  & $19.5 \pm 0.8 $& $16.6 \pm 0.7 $ \\
            $\rho^{0} \pi^0$ &  $6.2 \pm 0.6 $ & $6.48 \pm 0.25$ & $ 7.03 \pm 0.30$ \\
            $ \rho \eta$ & $0.193 \pm 0.023$ & $0.181 \pm 0.012 $ &$ 0.174 \pm 0.022 $ \\
            $\rho \eta'$ & $0.081 \pm 0.008$& $0.078 \pm 0.008$ & $ 0.082 \pm 0.008$
            \\
            $\omega \pi^{0}$ &$0.45 \pm 0.05$ & $0.47 \pm 0.04 $ &$ 0.48 \pm 0.04 $ \\
            $\phi\pi^0$
&
\begin{tabular}[c]{@{}l@{}}
$0.00294 \pm 0.00023$\\
$0.000124 \pm 0.000045$
\end{tabular}
&
$0.00205 \pm 0.00017$
&
$0.026 \pm 0.009$
\\
            \hline
    \end{tabular}
\caption{Experimental and theoretical branching ratios for VP decays of the $J/\psi$ with and without $g_{J,\mathrm{em}}$. The two experimental $\phi\pi^0$ values correspond to alternative interference solutions and are not included in the fit.}
    \label{tab: jpsi decays}
\end{table}

\newpage
\section{Discussion}
\subsection{Applicability of the $13\%$ rule}
The essence of the $\rho\pi$ puzzle is the breaking of the $13\%$ rule. In the previous section we have not used this rule to constrain the model, but we can 
see how well it applies to the result. To do this, we have to include phase space corrections to the $13\%$ rule consistent with the eLSM. We consider the following interaction for the decay of the $J/\psi$ (or $\psi(2S)$) into an electron-positron pair through a virtual photon, which has 2 interaction vertices, found in the Lagrangian terms given by
\begin{equation}
    \mathcal{L} = e A_{\mu} \bar{\psi} \gamma^{\mu} \psi - \frac{1}{2} \,g'_{V_{1} \gamma} V_{1,\mu\nu}F^{\mu\nu}.
\end{equation}
Here $\psi, \Bar{\psi}$ is the (anti)spinor of the electron (positron). Note that this $g'_{V_{1} \gamma}$ is not the same as the $g_{V_{1}\gamma}$ used in the previous section as that one includes averaging of nonperturbative effects of a high $q^2$ photon coupling to light hadrons. The resulting decay width depends on the phase space as
\begin{equation}
    \Gamma_{V_{1} e^{+}e^{-}} \propto \left(e\,g'_{V_{1}\gamma} \right)^2\frac{|\vec{k}|}{M_{V_{1}}^{2}}(M_{V_{1}}^2+2m_{e}^2),
    \label{eplus eminus decay}
\end{equation}
where we have suppressed some numerical prefactors, as they will be divided out later. By using the experimental result, we can now find the couplings $J/\psi -\gamma$ and $\psi(2S) -\gamma$ transitions, and therefore the ratio $r$ between $J/\psi$ and $\psi(2S)$ couplings, which is more or less the ratio between the wavefunctions at the origin in \eqref{waveft 0}. Defining $r$ as the ratio between the couplings so that
\begin{align}
g_{\psi} &= r\, g_{J/\psi} ,  
    \label{eq: psi2s coupling}
\end{align}
this gives the following result from the decays into electron-positron:
\begin{align}
        \frac{\mathcal{B}(\psi(2S) \to e^{+}e^{-})}{\mathcal{B}(J/\psi \to e^{+}e^{-})} \approx 13\% \Longrightarrow    r  \approx \pm 0.59.
        \label{eq: 13 percent rule coupling}
\end{align}
Since this ratio is meant to capture the ratio of the wavefunctions at zero, it should be universal among the couplings associated with a photon or 3-gluon channel. Note that, as $g_{JP}$ comes from a diagram where the $J/\psi$ does not decay, it would not be subject to this ratio if it were included in both fits.
\\
The ratios that come out of comparing the $\psi(2S)$ and $J/\psi$ fits do not match this ratio. Instead we find
\begin{equation}
    \begin{split}
\frac{g_\psi}{g_J} &=          0.092\pm 0.004, \\
\frac{g_{2\psi}}{g_{2J}} &=  0.037\pm 0.008, \\
\frac{g_{\psi\gamma}}{g_{J\gamma}} &=  -0.284\pm 0.025, \\
    \end{split}
\end{equation}
for the fit without $g_{J,\mathrm{em}}$, and 
\begin{equation}
    \begin{split}
\frac{g_\psi}{g_J} &=  0.106\pm 0.005, \\
\frac{g_{2\psi}}{g_{2J}} &=  0.125\pm 0.027, \\
\frac{g_{\psi\gamma}}{g_{J\gamma}} &= -0.37\pm 0.04, \\
    \end{split}
\end{equation}
when including $g_{J,\mathrm{em}}$. In both cases, the ratios are off by about a factor of 6 for the leading strong channels, a factor 6 or 16 for the higher order strong channel, and about a factor 2 for the coupling to the photon. We should keep in mind that the $g_{V_{1}\gamma}$ fit values include nonperturbative effects at different momenta of the virtual photon, at $q^{2}=M_{\psi(2S)}^2$ or $q^{2}=M_{J/\psi}^{2}$. As a result the comparison is not as direct as in the strong channels.
\subsection{The vector glueball}
\label{sec:glueball}
The existence of glueballs, bound states of gluons, is one of the oldest predictions of QCD, yet they have eluded experimental verification up to now. Lattice QCD has found a large spectrum of glueball states, with the vector glueball mass predicted to be in the range of 3.8 to 4 GeV \cite{Chen:2005mg,Athenodorou:2020ani} (for their direct comparison, see Ref. \cite{Trotti:2022knd}). 

In contrast to the lightest (pseudo)glueball states (see e.g. Refs. \cite{Janowski:2014ppa,Eshraim:2012jv} for eLSM predictions and Refs. \cite{Morningstar:2024vjk,Gross:2022hyw} for recent reviews), potential candidate states are lacking for the vector glueball. However, because of its shared quantum numbers with the $J/\psi$ and the $\psi(2S)$, the first experimental evidence of the vector glueball could be through its mixing. In addition, since the $J/\psi$ and $\psi(2S)$ decay into light hadrons primarily through a 3-gluon channel, their decays provide an especially interesting avenue for identifying glueballs experimentally. 
\\
\\
In the past a mixing between the vector glueball and the $J/\psi$ \cite{Brodsky:1987bb,Hou:1982kh}, as well as the $\psi(2S)$ \cite{Suzuki:2002bz} have been considered, but the situation is still not conclusive. There are two competing arguments for which of the two states has the largest mixing with the vector glueball. On the one hand, the glueball and the $J/\psi$ are both in their ground states, which presumably means the overlap is larger than between a ground state glueball and excited charmonium state. On the other hand, the mass of the $\psi(2S)$ is much closer to the recent quenched lattice predictions for the vector glueball mass, which lie in the range of about 3.8 to 4 GeV \cite{Chen:2005mg,Athenodorou:2020ani}. Unquenched lattice calculations are notoriously difficult and an unquenched prediction of the vector glueball mass is still missing \cite{Morningstar:2024vjk,Gregory:2012hu}. The unquenched mass might be low enough that the argument for $\psi(2S)$ mixing no longer holds. Moreover, our previous results show that the $\psi(2S)$ does not need any `hidden' coupling.
\\
\\
We could introduce a nonzero mixing of the $J/\psi$ with the vector glueball $O$. Calling the physical states $J/\psi$ as in the PDG listing, and $O'$, the pure charm-anticharm state $c\bar{c}$, and the pure glueball state $O$, the mixing is as follows:
\begin{equation}
        \begin{pmatrix} 
           J/\psi \\
           \mathcal{O}' \end{pmatrix} = \left( {\begin{array}{cc}
   \cos(\theta) & \sin(\theta) \\
   -\sin(\theta) & \cos(\theta) \\
  \end{array} } \right) \begin{pmatrix} 
        c\bar{c} \\
           ggg \equiv \mathcal{O} \end{pmatrix}.
           \label{mixing glueball}
\end{equation}
The mixing angle $\theta$ is assumed to be small, so $J/\psi$ is predominantly charm-anticharm and $O'$ is primarily gluonic. If one implements the mixing in this way for all interactions, we will have the new physical couplings
\begin{equation}
\begin{split}
 &g_{J}\to g'_{J}+g_{O} \\
    &g'_{J} \equiv g_{c\bar{c}}\cos(\theta), \,\,\,\,\, g_{O} \equiv g_{ggg}\sin(\theta).     
\end{split}
\end{equation}
However, when one does this, no new independent parameter is added. Neither $\theta$, $g'_{J}$ nor $g_{O}$ is separately measurable, only in the combination $g_{J}$. One way to resolve this is to demand the $12\%$ rule holds for the $g_{c\bar{c}}$ coupling and base this on the $\psi(2S)$ fit. This fixes $g'_{J}$ and leaves $g_{O}$ free. Imposing the $12\%$ rule only on the strong coupling leads to a fit equivalent to the one without $g_{J,\mathrm{em}}$, while also imposing it on $g_{J\gamma}$ (and assuming the direct coupling $g_{O\gamma}=0$) leads to a much worse fit, as the previous section suggests.
\\
Another option is to assume this substitution does not exactly hold for diagram (iii), that is, the process $O\to P\gamma \to PV$ does not go via the same VMD substitution as the same process for the $J/\psi$ does. This does not require a direct $O-\gamma$ coupling, since here $O$ couples to $\gamma$ through the light vector mesons. This will be equivalent to the fit we have with $g_{J,\mathrm{em}}$, but with the interpretation that the VMD substitution is not exact for the glueball rather than for the $J/\psi$.
\\
Assuming that the glueball does not enter diagram (iii) at all leads to a very large cancellation between $g_{J}'$ and $g_{O}$ so that the coefficient for the leading strong diagram is $g_{J}'+g_{O} = (5.91\pm0.17)\times 10^{-7}$. Such a peculiar cancellation may be rightly regarded as a drawback, but this can be relaxed as the equivalence holds even if the glueball does enter diagram (iii), as long as it is not with the same coefficient as the $J/\psi$. These arguments do not mean that the results should be interpreted for or against mixing with a vector glueball. Rather, in these decays we cannot distinguish whether glueball mixing or some other dynamics of the $J/\psi$ is responsible.
\\
\\
In order to identify the vector glueball in future studies, it would be useful to know approximately its width. While a reliable lattice QCD determination of the vector glueball width is presently unavailable, we can use large $N_{c}$ arguments for an estimate. Following \cite{Giacosa:2017eqy}, OZI-allowed decays of conventional mesons scale as $N_{c}^{-1}$, glueball decays into mesons as $N_{c}^{-2}$, and OZI-suppressed decays as $N_{c}^{-3}$. We would then expect the vector glueball decay width to be between the OZI-suppressed and OZI-allowed decays of the vector states. For comparison, the $J/\psi$ has a total decay width of order $ 0.1$ MeV and the  $\psi(4040)$ has a total decay width of order $100$ MeV. Based on this, \cite{Giacosa:2017eqy} proposed the rough estimate 
\begin{equation}
    \Gamma_{\mathcal{O}}\sim 10 \text{ MeV},
\end{equation}
which should be understood only as a heuristic order of magnitude expectation. In fact, a glueball significantly broader than this expectation has come up in recent literature. In a holographic approach \cite{Hechenberger:2024piy} the vector glueball width was found to be quite large. In a similar holographic calculation for the pseudovector glueball \cite{Brunner:2018wbv}, the leading two-body decay indeed scales as $N_{c}^{-2}$, but still gives a very broad decay width when physical values are inserted. For the tensor glueball, holographic \cite{Brunner:2015oqa} and eLSM \cite{Vereijken:2023jor} calculations both suggest a broad tensor glueball.

\section{Conclusion}
In this work we have studied the $VP$ decays of the $J/\psi$ and $\psi(2S)$ within an effective chiral framework including both strong and electromagnetic interactions. The $\psi(2S)$ is described well by a relatively minimal construction with three free parameters, giving a reduced chi-squared of $\chi^{2}_{\text{red}}\sim 1.12$, and we predict the branching ratio of some not-yet-measured decay channels. Although the $12\%$ rule is broken, the $\psi(2S)$ data are well described without introducing any additional mechanism beyond the strong and electromagnetic contributions of the minimal model.
\\
On the other hand, the $J/\psi$ decays are not easily described. The minimal description for strong and electromagnetic contributions is not enough to describe the data. We require further $\mathrm{SU}(3)$ flavor breaking and electromagnetic contributions, finally reaching an acceptable $\chi^{2}_{\text{red}}\sim 1.97$. 
\\
The fit with additional electromagnetic contributions is equivalent to a fit where a mixing with the vector glueball is included under certain assumptions. Thus, the eventual presence of the latter state within hadronic decays of the $J/\psi$ cannot be singled out.
\\
We also compare the values of fitted couplings to the expectation of the $12\%$ rule, and find that this rule is still substantially violated in every description, making an adjustment of the fit where the $12\%$ rule is implemented difficult.
Taking these analyses together, our results suggest that the solution to the $\rho\pi$ puzzle is likely to be an unexpected enhancement of $J/\psi\to VP$ decays, rather than one of $\psi(2S)$ suppression.
\\
\\

\textbf{Acknowledgments}: A. V. thanks A. Rebhan and F. Hechenberger for useful discussions. 

\bigskip

\textbf{AI disclosure:}  ChatGPT (OpenAI) was used for manuscript review, numerical cross-checks, and refinement of the text. The model, calculations, numerical fits, results, and conclusions are those of the authors.

\clearpage

\appendix

\section{Mesonic nonets}
\label{appendix:nonets}
The scalar chiral nonet consists of the scalar and pseudoscalar nonets as
\begin{equation}
    \Phi = S + iP,
\end{equation}
and transforms under $\mathrm{U}(3)_L\times \mathrm{U}(3)_R$ chiral symmetry as 
\begin{equation}
    \Phi \to  U_{L}\Phi U_{R}^{\dagger}.
\end{equation}
The ground state pseudoscalar mesons $P$ with $J^{PC}=0^{-+}$ are the 3 pions, four kaons, the $\eta(547)$ and the $\eta^\prime(958)$. The meson matrix (nonet) is given by
\begin{equation}
P=\frac{1}{\sqrt{2}}%
\begin{pmatrix}
\frac{\eta_{N}+\pi^{0}}{\sqrt{2}} & \pi^{+} & K^{+}\\
\pi^{-} & \frac{\eta_{N}-\pi^{0}}{\sqrt{2}} & K^{0}\\
K^{-} & \bar{K}^{0} & \eta_{S}%
\end{pmatrix}
\,\text{,}
\label{eq:pseudoscalar_nonet}%
\end{equation}
where the $\eta_{N}$ and $\eta_{S}$ are made up of nonstrange and strange quarks respectively: 
\begin{equation}
    \eta_{N}\sim\frac{1}{\sqrt{2}}\,(\bar{u}i\gamma^{5}u+\bar{d}i\gamma^{5}d), \ \, \eta_{S} \sim \bar{s}i\gamma^{5}s.
\end{equation}
The physical fields are a mixture of these states:
\begin{equation}
\left(
\begin{array}
[c]{c}%
\eta\\
\eta'
\end{array}
\right)  =\left(
\begin{array}
[c]{cc}%
\cos\beta_{P} & \sin\beta_{P}\\
-\sin\beta_{P} & \cos\beta_{P}%
\end{array}
\right)  \left(
\begin{array}
[c]{c}%
\eta_{N}\\
\eta_{S}%
\end{array}
\right)  \text{ ,}
\end{equation}
with a mixing angle $\beta_{P}=-43.4^{\circ}$ \cite{Kloe2}. The large mixing angle is a consequence of the $U_{A}(1)$ axial anomaly \cite{Feldmann:1998vh,tHooft:1986ooh}. 
\\
The scalar chiral nonet has a nonzero vev due to spontaneous chiral symmetry breaking. We denote $\Phi \to \Phi + \Phi_0$, with the condensate $\Phi_0$ given as
\begin{equation}
    \Phi_0=\frac{1}{\sqrt{2}}\begin{pmatrix}
\frac{\phi_N}{\sqrt{2}} & 0 &
0\\
0 & \frac{\phi_N}{\sqrt{2}} & 0\\
0 & 0 & \phi_S%
\end{pmatrix}
\end{equation}
where the nonstrange and strange condensates respectively have the numerical values $\phi_{N}=0.158$ GeV and $\phi_{S}=0.138$ GeV \cite{Jafarzade:2022uqo,Parganlija:2012fy,Giacosa:2024epf}.
\\
\\
The ground state vector mesons are the states \{$\rho(770),$ $K^{\ast}(892),$ $\omega(782),$ $\phi(1020)$\}. The matrix $V^{\mu}$ has the form
\begin{equation}
V^{\mu}=\frac{1}{\sqrt{2}}%
\begin{pmatrix}
\frac{\omega_{N}^{\mu}+\rho^{0\mu}}{\sqrt{2}} & \rho^{+\mu} &
K^{\ast+\mu}\\
\rho^{-\mu} & \frac{\omega_{N}^{\mu}-\rho^{0\mu}}{\sqrt{2}} & K^{\ast0\mu}\\
K^{\ast-\mu} & \bar{K}^{\ast0\mu} & \omega_{S}^{\mu}%
\end{pmatrix}
\,, \label{eq:vector_nonet}%
\end{equation}
where the mixing of the non-strange and strange states $\omega_{N}$ and $\omega_{S}$ is similar to the pseudoscalar case:
\begin{equation}
\left(
\begin{array}
[c]{c}%
\omega(782)\\
\phi(1020)
\end{array}
\right)  =\left(
\begin{array}
[c]{cc}%
\cos\beta_{V} & \sin\beta_{V}\\
-\sin\beta_{V} & \cos\beta_{V}%
\end{array}
\right)  \left(
\begin{array}
[c]{c}%
\omega_{N}\\
\omega_{S}%
\end{array}
\right)  \text{ ,}%
\end{equation}
where the small vector mixing angle $\beta_{V}=-3.9^{\circ}$ \cite{ParticleDataGroup:2024cfk}. In contrast to the pseudoscalar sector, this mixing angle is small, so $\omega$ is predominantly nonstrange while $\phi$ is predominantly of strange quarks. 
\\
\\
When constructing chiral Lagrangians, it is useful to define the left- and right-handed fields
\begin{align}
    R^{\mu} &= V^{\mu}-A_{1}^{\mu}, \\
    L^{\mu}&=V^{\mu}+A_{1}^{\mu},
\end{align}
where $A_{1}^{\mu}$ are the axial vector mesons. These have a simple transformation law under chiral symmetry:
\begin{equation}
       R^{\mu} \to U_{R}R^{\mu}U_{R}^{\dagger}, \, \, L^{\mu}\to U_{L}L^{\mu}U_{L}^{\dagger}.
\end{equation}
Finally, due to spontaneous breaking of chiral symmetry, the pseudoscalars and axial vector mesons acquire a mixing. This is implemented by the shift $A_{1}^{\mu} \to A_{1}^{\mu} + \partial^{\mu}\mathcal{P}$, where $\mathcal{P}$ is a renormalized pseudoscalar nonet given by
\begin{equation}
    \mathcal{P} = \frac{1}{\sqrt{2}}
		\begin{pmatrix}
			\frac{Z_{\pi}w_{\pi}(\eta_N + \pi^0)}{\sqrt{2}} & Z_{\pi}w_{\pi}\pi^+ & Z_Kw_K K^+ \\
			Z_{\pi}w_{\pi}\pi^- & \frac{Z_{\pi}w_{\pi}(\eta_N - \pi^0)}{\sqrt{2}} &  Z_Kw_K K^0 \\
			 Z_Kw_K K^- &  Z_Kw_K \bar{K}^0 & Z_{\eta_S}w_{\eta_S}\eta_S
		\end{pmatrix}.
\end{equation}
The factors $Z_\pi, Z_K,Z_{\eta_S},Z_{\eta_N}$ are wavefunction renormalization constants and as such will also appear in the ordinary pseudoscalar nonet $P$ after chiral symmetry breaking. The $\omega's$ are due to the mixing and only appear with $A_{1}$ (or $\mathcal{P}$). The numerical values for these constants are \cite{Giacosa:2024epf,Parganlija:2012fy}
\begin{align}
    Z_{\pi}&= 1.709, \\
    Z_K &= 1.604, \\
    Z_{\eta_S} &= 1.539, \\
    \omega_{\pi}&= 0.683 \text{ GeV}^{-1}, \\
    \omega_{K} &= 0.611 \text{ GeV}^{-1}, \\
    \omega_{\eta_{S}} &= 0.554 \text{ GeV}^{-1}.
\end{align}

\clearpage
\renewcommand\bibname{Bibliography}
\bibliographystyle{utphys.bst}
\bibliography{bibliography}

\end{document}